\documentclass[%
 reprint,
twocolumn, 
 amsmath,amssymb,
 aps,
prl,
superscriptaddress
]{revtex4-2}

\usepackage{graphicx}
\usepackage{dcolumn}
\usepackage{bm}
\usepackage[colorlinks=true, allcolors=blue,breaklinks]{hyperref}
\usepackage{braket}
\usepackage{bbding}
\usepackage{natbib}
\usepackage{gensymb}
\usepackage{balance}
\usepackage{pifont}
\usepackage{longtable}
\usepackage{microtype}

\usepackage{amsmath,amssymb,amsfonts,mathrsfs}
\usepackage{CJK}
\usepackage{bm}
\usepackage{booktabs}
\usepackage{multirow}
\usepackage{longtable}
\usepackage{epstopdf}
\usepackage{dcolumn} 
\usepackage{babel}
\usepackage{graphicx}
\usepackage{float}
\allowdisplaybreaks
\usepackage{braket}
\usepackage{color}
\usepackage{makecell}
\usepackage{bbding}
\usepackage{utfsym}
\usepackage{pifont}
\usepackage{graphicx}
\usepackage{dcolumn}
\usepackage{bm}
\usepackage[colorlinks=true, allcolors=blue]{hyperref} 

\begin{document}


\title{Large Pockels Effect in Multiaxial Ferroelectrics Induced by Polarization Reorientation}

\author{Zhijun Jiang}
\affiliation{%
 Ministry of Education Key Laboratory for Nonequilibrium Synthesis and Modulation of Condensed Matter, Shaanxi Province Key Laboratory of Advanced Functional Materials and Mesoscopic Physics, School of Physics, Xi’an Jiaotong University, Xi’an 710049, China 
}%
\affiliation{%
 Key Laboratory of Material Simulation Methods and Software of Ministry of Education,
College of Physics, Jilin University, Changchun 130012, China
}%

\author{Jinyang Ni}
\affiliation{%
 Ministry of Education Key Laboratory for Nonequilibrium Synthesis and Modulation of Condensed Matter, Shaanxi Province Key Laboratory of Advanced Functional Materials and Mesoscopic Physics, School of Physics, Xi’an Jiaotong University, Xi’an 710049, China 
}%

\author{Hong Jian Zhao}
\email{physzhaohj@jlu.edu.cn}
\affiliation{%
 Key Laboratory of Material Simulation Methods and Software of Ministry of Education,
College of Physics, Jilin University, Changchun 130012, China
}%
\affiliation{%
 Key Laboratory of Physics and Technology for Advanced Batteries (Ministry of Education), College of Physics, Jilin University, Changchun 130012, China
}%
\affiliation{%
International Center of Future Science, Jilin University, Changchun 130012, China
}%


\begin{abstract}
Materials with large linear electro-optic (Pockels) effect have significant potential for fabricating energy-efficient photonic devices. Previous strategies have mostly relied on structural softness---associated with phase boundary, domain wall, or strain induced phase transition---to create large Pockels effect in ferroelectrics. In this Letter, we show that polarization reorientation in multiaxial ferroelectrics may create structural softness and thus induce large Pockels effect. By developing a phenomenological theory, we find that multiaxial ferroelectrics exhibit a hysteretic relationship between the azimuth angles of the polarization and the electric field, and a possible divergence of the linear electro-optic coefficient may occur at an appropriate polarization azimuth angle. This suggests a brand-new strategy for engineering large Pockels effect in multiaxial ferroelectrics. We perform first-principles simulations on several multiaxial ferroelectrics and predict that polarization reorientation can significantly enhance the electro-optic coefficients in monolayer SnSe and bulk KNbO$_3$. This validates the soundness of our strategy and emphasizes multiaxial ferroelectrics as promising platforms toward large Pockels effect.
\end{abstract}

\maketitle
\newpage

\textit{Introduction—}The Pockels effect is referred to as a change of the refractive index induced by an electric field and is characterized by the linear electro-optic coefficient \cite{Li2019,Hu2025,Wang2025,Boes2023,Thomaschewski2022}. Materials with large Pockels effect are of great promise for developing energy-efficient photonic devices such as 
electro-optic modulators~\cite{Wang2018,He2019,Eltes2020}, electro-optic switches~\cite{Thomaschewski2020,Liu2022}, and microwave-to-optical quantum transducers~\cite{abel2019,Hease2020,Sahu2023,Youssefi2021,ulrich2025}. Ferroelectrics are potential large Pockels materials in which the induced structural softness enables sizable electro-optic responses \cite{Wemple1968,DiDomenico1969,Zgonik1994,veithen2004,Paillard2019,Jiang2019,jiang2020,Paoletta2021,jiang2022}. It is shown that phase-boundary engineering~\cite{Barad2000,Deng2026,Chen2026}, domain engineering~\cite{Qiu2020,Zhu2010,Liu2022,Shang2022,Li2022,Chen2025}, and strain engineering~\cite{Hamze2018,Fredrickson2018,Paillard2019,Jiang2019,Delodovici2023,zhang2024Giant,Wan2025,Suceava2026} are effective strategies for tailoring structural softness and achieving large Pockels effect in ferroelectrics. In this regard, phase-boundary engineering enables a relatively flat free-energy landscape formed by nearly degenerate ferroelectric phases~\cite{Barad2000,Deng2026,Chen2026}, domain engineering enables a coexistence of different polarization states near the domain walls~\cite{Qiu2020,Zhu2010,Shang2022,Li2022,Chen2025}, and strain engineering softens the phonon modes by driving structural phase transitions~\cite{Hamze2018,Fredrickson2018,Paillard2019,Jiang2019,Delodovici2023,zhang2024Giant,Wan2025,Suceava2026}. Overall, these strategies reduce the energetic cost for polarization change by taking advantage of the structural softness in the vicinity of, e.g., morphotropic phase boundary, domain wall, or critical strain.

In this Letter, we propose a brand-new strategy for engineering large Pockels effect, which harnesses the polarization reorientation induced structural softness in multiaxial ferroelectrics. We develop a phenomenological theory describing the electric field induced polarization reorientation and the associated linear electro-optic response. Our theory suggests that (i) the orientation of the polarization (characterized by $\phi$) can be manipulated by an oriented electric field (characterized by $\theta$), (ii) the $\phi$ and $\theta$ azimuth angles exhibit a hysteresis relationship resembling to the ferroelectric hysteresis loop, and (iii) the electro-optic response may present a divergent behavior at a critical $\phi$ angle (yielding a large coefficient). Then, we verify our theory by working with SnS, SnSe, BaTiO$_3$, and KNbO$_3$, which are two-dimensional or bulk multiaxial ferroelectrics. Our first-principles simulations show that large linear electro-optic coefficients (i.e., $\sim$436 pm/V in SnSe and $\sim$226 pm/V in KNbO$_3$) are enabled by engineering the polarization orientation $\phi$. \\

\begin{figure*}
    	\centering
   	\includegraphics[width=1\textwidth]{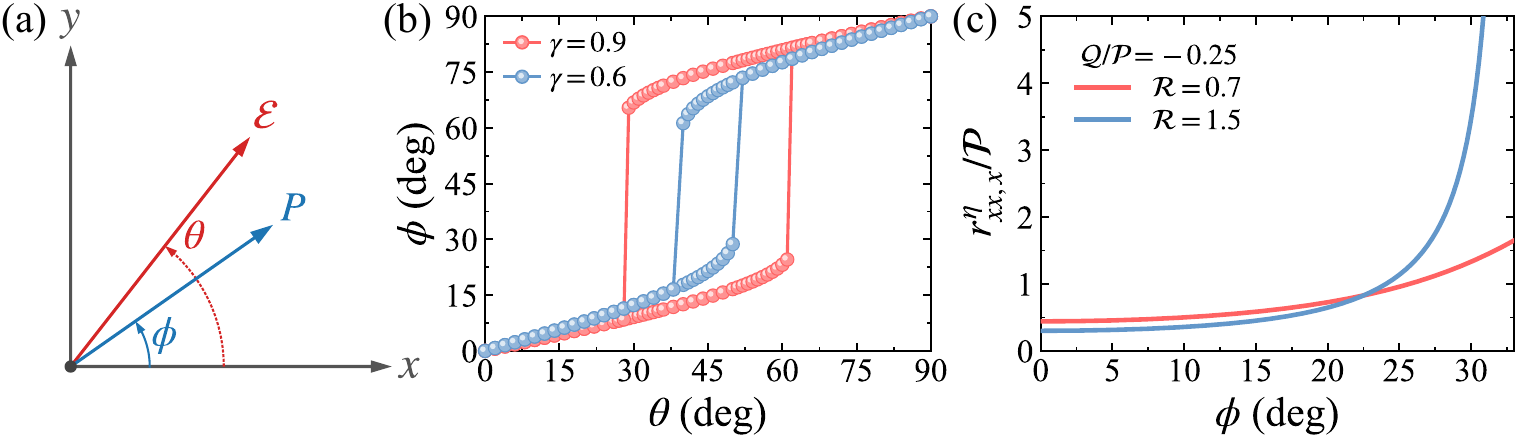}
    \caption{The polarization reorientation and the electro-optic effect in multiaxial ferroelectrics. Panel (a) sketches the orientations of the electric field and electric polarization, where $\mathcal{E}$ and $P$ represent their magnitudes. In the polar coordinate system, $\mathcal{E}_x$, $\mathcal{E}_y$, $P_x$, and $P_y$ are given by $\mathcal{E}\mathrm{cos}(\theta)$, $\mathcal{E}\mathrm{sin}(\theta)$, $P\mathrm{cos}(\phi)$, and $P\mathrm{sin}(\phi)$, respectively. Panel (b) displays the $\phi$-$\theta$ hysteresis loops under the electric field of $\mathcal{E}=\frac{3\alpha^{3/2}}{8\sqrt{2\beta}}$. Panel (c) shows the dependence of the $r^\eta_{xx,x}$ coefficient on the $\phi$ azimuth angle.}
    \label{fig1}
\end{figure*}

\textit{The $\phi$-$\theta$ hysteresis loop—}First of all, we establish a theory that describes the $\phi$-$\theta$ hysteresis behavior in multiaxial ferroelectrics. 
To simplify our discussion, we work with a minimal energetic model with the free energy $\mathcal{U}$ given by
\begin{equation}\label{eq:freeenergy}
\begin{aligned}
\mathcal{U} = {} & - \alpha(P_x^2 + P_y^2) + \beta(P_x^4 + P_y^4) \\
& + (\gamma+2)\beta P_x^2 P_y^2 - \mathcal{E}_x P_x - \mathcal{E}_y P_y,
\end{aligned}
\end{equation}
where $P_i$ is the polarization, $\mathcal{E}_i$ is the electric field, $i=x,y$ labels the Cartesian orientation of the polarization or the electric field, and the $\alpha,\beta,\gamma$ coefficients are assumed to be positive. In particular, the $(\gamma+2)\beta P_x^2 P_y^2$ term describes the polarization anisotropy in multiaxial ferroelectrics. Under null electric field (i.e., $\mathcal{E}_x=\mathcal{E}_y=0$), the spontaneous polarization is along $\pm x$ or $\pm y$ direction with a magnitude of $\sqrt{\alpha/2\beta}$. In the polar coordinate system [Fig.~\ref{fig1}(a)], Eq.~(\ref{eq:freeenergy}) is written as 
\begin{equation}\label{eq:freeeergyrw}
\begin{aligned}
\mathcal{U} = -\alpha P^2 + \beta \left[ 1 +  \frac{\gamma}{4}  \sin^2(2\phi) \right] P^4  - \mathcal{E} P \cos(\theta - \phi).
\end{aligned}
\end{equation}
By setting $P\approx\sqrt{\alpha/2\beta}$, Eq.~(\ref{eq:freeeergyrw}) is  approximated as 
\begin{equation}\label{eq:freeenergy2}
\begin{aligned}
\mathcal{U} \approx  - \frac{\alpha^2}{2\beta} + \frac{\alpha^2}{4\beta} \left[ 1 + \frac{\gamma}{4} \sin^2(2\phi) \right] - \sqrt{\frac{\alpha}{2\beta}} \mathcal{E} \cos(\theta - \phi).
\end{aligned}
\end{equation}
Under a $(\mathcal{E},\theta)$ electric field, the thermodynamically allowed $\phi$ azimuth angles satisfy
\begin{align}\label{eq:condition1}
\partial{\mathcal{U}(\phi)}/\partial{\phi}=0 \implies \mathrm{sin}(\theta-\phi)=\frac{\alpha^{3/2}\gamma}{4\sqrt{2\beta}\mathcal{E}}\mathrm{sin}(4\phi),
\end{align}
and
\begin{align}\label{eq:condition2}
\partial^2{\mathcal{U}(\phi)}/\partial{\phi^2}>0.
\end{align}
As shown in Fig.~\ref{fig1}(b), the numerical solution of Eqs.~(\ref{eq:condition1}) and~(\ref{eq:condition2}) indicates the hysteresis loop of $\phi$ versus $\theta$. In particular, the anisotropic $\gamma$ coefficient determines the horizontal width of the hysteresis loop.\\

\textit{The divergence of the electro-optic response—}We move on to explore the relationship between the electro-optic coefficient and the azimuth angle of the
polarization. The Hessian matrix $\mathcal{X}$ and the electric susceptibility $\chi$ are defined as 
\begin{equation}\label{eq:hessian}
\begin{aligned}
\mathcal{X}\equiv 
\begin{pmatrix}
\mathcal{X}_{xx}  & \mathcal{X}_{xy} \\
\mathcal{X}_{yx} & \mathcal{X}_{yy}
\end{pmatrix}
\end{aligned},
\begin{aligned}
\chi\equiv 
\begin{pmatrix}
\chi_{xx}  & \chi_{xy} \\
\chi_{yx} & \chi_{yy}
\end{pmatrix}
\end{aligned},
\end{equation}
where $\mathcal{X}_{xx}=-2\alpha+12\beta P_x^2+2(\gamma+2)\beta P_y^2$, $\mathcal{X}_{yy}=-2\alpha+12\beta P_y^2+2(\gamma+2)\beta P_x^2$, and $\mathcal{X}_{xy}=\mathcal{X}_{yx}=4(\gamma+2)\beta P_x P_y$ are derived from Eq.~(\ref{eq:freeenergy}). Furthermore, the $\chi=\mathcal{X}^{-1}$ relationship is suggested by Ref.~\cite{softmode} and derived in Section I of the Supplementary Material \cite{SM}.
In the large-susceptibility limit, the dielectric permittivity tensor $\epsilon$ is approximated by $\epsilon\approx\chi/\epsilon_0$ ($\epsilon_0$ being the vacuum dielectric permittivity). The Pockels $r^\eta_{ij,k}$ electro-optic coefficient is obtained by $r^\eta_{ij,k}=\partial (\epsilon^{-1})_{ij}/\partial{\varepsilon_k} \approx \epsilon_0 \partial (\chi^{-1})_{ij}/\partial{\varepsilon_k}=\epsilon_0 \partial \mathcal{X}_{ij}/\partial{\varepsilon_k}$, where $\varepsilon_k$ is the electric field for changing the dielectric permittivity (refractive index), and $i,j,k=x,y$ label the Cartesian orientation \cite{veithen2005,veithen2005temperature,prosandeev2024}. As a demonstration, we derive the $r^\eta_{xx,x}$ electro-optic coefficient as a function of the $\phi$ azimuth angle. Working with Eq.~(\ref{eq:hessian}), the $r^\eta_{xx,x}$ coefficient is given by
\begin{align}\label{eq:rxxx}
r^\eta_{xx,x}=\epsilon_0 \left( \frac{\partial{\mathcal{X}_{xx}}}{\partial{P_x}}\chi_{xx} + \frac{\partial{\mathcal{X}_{xx}}}{\partial{P_y}}\chi_{yx} \right ).
\end{align}

Using the $P_x\approx\sqrt{\alpha/2\beta}\mathrm{cos}(\phi)$ and $P_y\approx\sqrt{\alpha/2\beta}\mathrm{sin}(\phi)$ approximation, Eq.~(\ref{eq:rxxx}) can be further written as
\begin{align}\label{eq:rxxx2}
r^\eta_{xx,x}=\frac{\tau[(8+5\gamma-\gamma^2)\mathrm{cos}(\phi)+(-8+7\gamma+\gamma^2)\mathrm{cos}(3\phi)]}{\gamma[8-3\gamma+(24+3\gamma)\mathrm{cos}(4\phi)]},
\end{align}
with $\tau\equiv 8\epsilon_0 \sqrt{2\beta/\alpha}$. Upon introducing a $\zeta (P_x^6+P_y^6)$ correction to Eq.~(\ref{eq:freeenergy}), a more general form of Eq.~(\ref{eq:rxxx}) can be derived:
\begin{equation}
\label{eq:rxxx_fit}
r^\eta_{xx,x}=
\frac{
\mathcal{P}\cos\phi+\mathcal{Q}\cos(3\phi)
}{
1+\mathcal{R}\cos(4\phi)},
\end{equation}
where $\mathcal{P}$, $\mathcal{Q}$, and $\mathcal{R}$ are parameters to be fitted (see Section I of the Supplementary Material \cite{SM}). According to Eq.~(\ref{eq:rxxx_fit}), the $r^\eta_{xx,x}$ coefficient depends on the $\phi$ angle; if $|\mathcal{R}|\ge1$, $r^\eta_{xx,x}$ becomes divergent at $\phi=\mathrm{arccos}(-1/\mathcal{R})/4$ [see Fig.~\ref{fig1}(c)]. This implies that the reorientation of polarization may create large electro-optic responses in multiaxial ferroelectrics. \\

\begin{figure}
    	\centering
   	\includegraphics[width=0.5\textwidth]{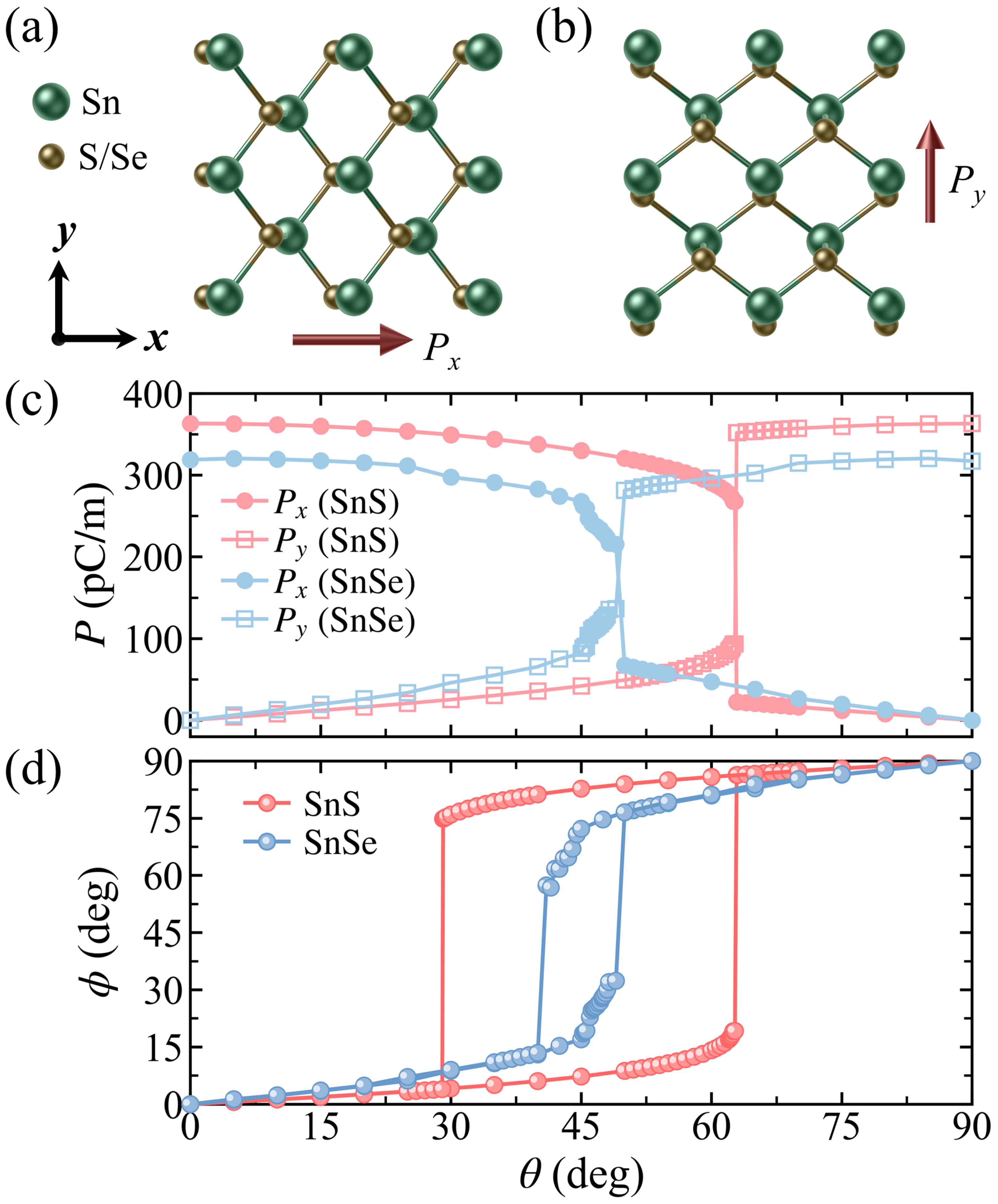}
    \caption{The crystal structures and hysteresis loops for SnS and SnSe. Panels (a) and (b) sketch the atomic displacements for Sn$X$ ($X$ being S or Se) with $P_x$ and $P_y$ polarization states, respectively. Panel (c) shows the \(P_x\) and \(P_y\) polarization components as functions of the orientation of the electric field. Panel (d) displays the $\phi$-$\theta$ hysteresis loops. The magnitude of the electric field is $2.5~\mathrm{MV/cm}$. See Fig.~\ref{fig1}(a) for the definitions of $\theta$ and $\phi$ azimuth angles.}
    \label{fig2}
\end{figure}

\textit{Pockels effect in two-dimensional ferroelectrics—}In the following, we shall validate our theory and verify the effectiveness of our strategy. Our model involves $P_x$ and $P_y$ polarizations and is tailor-made for describing the polarization reorientation and the electro-optic effect in two-dimensional ferroelectrics. In view of this, we select monolayer SnS and SnSe ferroelectric materials as our test beds. According to experiments~\cite{higashitarumizu2020,chang2020}, these materials are room-temperature biaxial ferroelectrics, and their crystal structures are similar to each other. The $P_x$ and $P_y$ polarization states for SnS and SnSe are sketched in Figs.~\ref{fig2}(a) and~\ref{fig2}(b). The space groups associated with $P_x$ and $P_y$ states are $P2_1 mn$ and $Pm 2_1 n$, respectively~\footnote{Note that $P2_1 mn$ and $Pm 2_1 n$ are non-standard settings of the $Pmn2_1$ space group.}. We investigate the ferroelectric and electro-optic behaviors of SnS and SnSe by first-principles simulations. The computational details are shown in Section II of the Supplementary Material \cite{SM}.

Figure~\ref{fig2}(c) shows the $P_x$ and $P_y$ polarization values (for SnS and SnSe) as functions of the $\theta$ azimuth angle. The $\theta$ angle defines the orientation of the applied electric field, where $\theta=0^{\circ}$ and $\theta=90^{\circ}$ correspond to electric fields along $x$ and $y$ directions, respectively [see Fig.~\ref{fig1}(a)]. Upon increasing $\theta$, the $P_x$ ($P_y$) polarization of SnS decreases (increases) slightly and then drops (rises) sharply at $\theta \approx 63^\circ$. The situation of SnSe is similar to that of SnS, but with a different critical $\theta$ angle (i.e., $\theta\approx 49^\circ$) for the abrupt changes of the polarization. The $\phi$-$\theta$ hysteresis loops for monolayer SnS and SnSe are shown in Fig.~\ref{fig2}(d); the results are in excellent agreement with our theory [see Eq.~(\ref{eq:condition1}), Eq.~(\ref{eq:condition2}), and Fig.~\ref{fig1}(b)].

\begin{figure}
    	\centering
   	\includegraphics[width=0.5\textwidth]{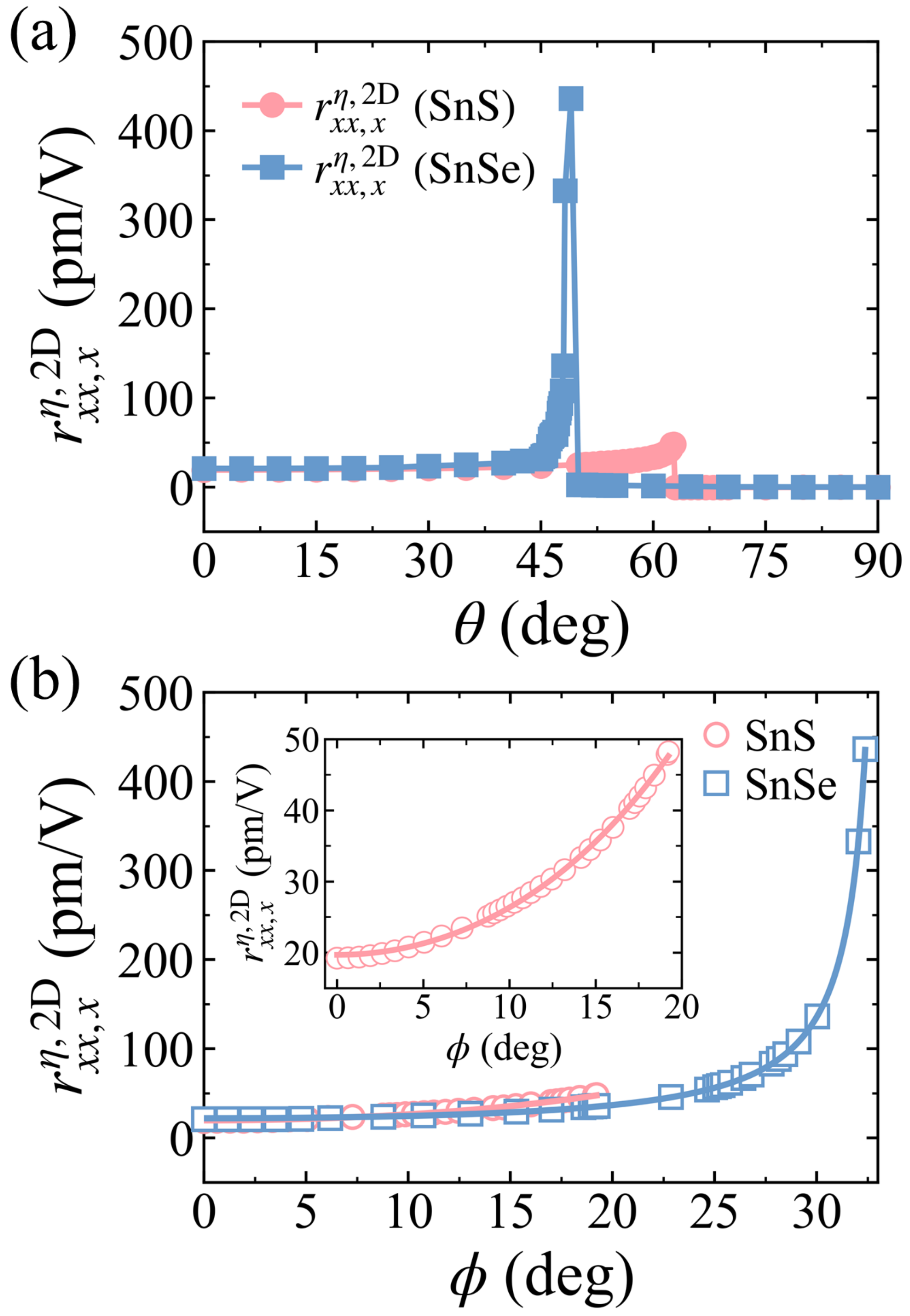}
    \caption{The Pockels effect in SnS and SnSe. Panels (a) and (b) show the \(r^{\eta,\mathrm{2D}}_{xx,x}\) coefficients as functions of \(\theta\) and \(\phi\) azimuth angles, respectively. The subscript ``2D'' indicates that the electro-optic coefficients are defined with respect to two-dimensional materials. In panel (b), the \(r^{\eta,\mathrm{2D}}_{xx,x}\)-\(\phi\) relationships are fitted with Eq.~(\ref{eq:rxxx_fit}), where the fitting results are shown by the blue and pink curves. The fitting \((\mathcal{P},\mathcal{Q},\mathcal{R},R^2)\) results are: \((91.98,-61.81,0.48, 0.99)\) for SnS and \((43.54,10.22,1.44, 0.99)\) for SnSe, where $R^2$ represents the coefficient of determination, \(\mathcal{P}\) and \(\mathcal{Q}\) are in units of pm/V, and \(\mathcal{R}\) is dimensionless.}
    \label{fig3}
\end{figure}

\begin{figure*}
    	\centering
   	\includegraphics[width=1\textwidth]{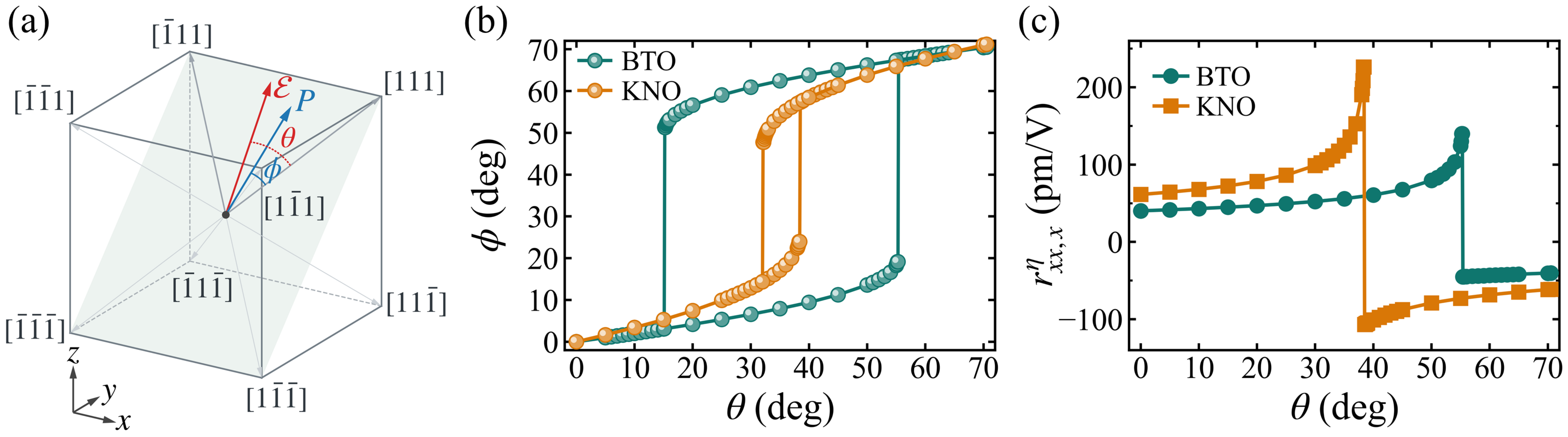}
    \caption{The Pockels effect in BaTiO$_3$ and KNbO$_3$. Panel (a) depicts the eight symmetrically equivalent polarization states. Each $[uvw]$ crystallographic direction index ($u,v,w$ being $1$ or $\bar{1}$) labels a specific polarization state. Panels (b) and (c) show the $\theta$-dependent hysteresis loops and the electro-optic coefficients, where BaTiO$_3$ and KNbO$_3$ are abbreviated as BTO and KNO, respectively. The magnitude of the electric field is fixed at $0.3~\mathrm{MV/cm}$}
    \label{fig4}
\end{figure*}

The clamped electro-optic coefficients for SnS and SnSe are calculated by density-functional perturbation theory, with both electronic and ionic contributions. In the absence of external electric field, the \(r^{\eta,\mathrm{2D}}_{xx,x}\) coefficients for SnS and SnSe are $\sim$33 and $\sim$57 pm/V, respectively. Such electro-optic coefficients become $\sim$19 for SnS and $\sim$21 for SnSe, when applying an electric field of 2.5 MV/cm along their polarization directions (i.e., $\theta=0^\circ$). 
As shown in Fig.~\ref{fig3}(a), the \(r^{\eta,\mathrm{2D}}_{xx,x}\) coefficients for SnS and SnSe are tunable by varying the azimuth angle of the electric field. The divergent behavior of \(r^{\eta,\mathrm{2D}}_{xx,x}\) is observed in SnSe, with a sharp increase from $\sim$136 pm/V at $\theta\approx48^\circ$ to $\sim$436 pm/V at $\theta\approx49^\circ$.  
In contrast, changing the $\theta$ angle slightly changes the \(r^{\eta,\mathrm{2D}}_{xx,x}\) coefficient of SnS: from $\sim$40 pm/V at $\theta\approx62^\circ$ to $\sim$48 pm/V at $\theta\approx63^\circ$. Figure~\ref{fig3}(b) shows the \(r^{\eta,\mathrm{2D}}_{xx,x}\) coefficients as functions of $\phi$. For both SnS and SnSe, the relationships between \(r^{\eta,\mathrm{2D}}_{xx,x}\) and \(\phi\) are well fitted by Eq.~(\ref{eq:rxxx_fit}), which demonstrates the robustness of our theory.\\

\textit{Pockels effect in bulk ferroelectrics—}The enhancement of the Pockels response (by polarization reorientation) is not unique to two-dimensional biaxial ferroelectrics. Such a phenomenon can occur in bulk multiaxial ferroelectrics that go beyond the scope of Eqs.~(\ref{eq:condition1}),~(\ref{eq:condition2}), and~(\ref{eq:rxxx_fit}). To demonstrate this, we work with ferroelectric BaTiO$_3$ and KNbO$_3$, both of which have a ferroelectric rhombohedral $R3m$ phase and a paraelectric cubic $Pm\bar{3}m$ phase~\cite{hewat1973BTO,hewat1973KNO,king1994,zhong1994,zhong1995,krakauer1999}. As shown in Fig.~\ref{fig4}(a), rhombohedral BaTiO$_3$ and KNbO$_3$ have eight symmetrically equivalent polarization states. The free energy for these two materials is given by $\mathcal{U}=-\alpha (P_x^2+P_y^2+P_z^2) + \beta(P_x^4+P_y^4+P_z^4)+\zeta (P_x^6 + P_y^6 + P_z^6) + \gamma (P_x^2 P_y^2 + P_y^2 P_z^2 + P_z^2 P_x^2) + \kappa P_x^2 P_y^2 P_z^2+\cdots$, where $P_x, P_y, P_z$ are polarization components and $\alpha,\beta,\zeta,\gamma,\kappa$ the coefficients. The compact analytical solution for such a complicated model is difficult to obtain. Yet, we can still investigate the $\theta$-dependent Pockels responses in BaTiO$_3$ and KNbO$_3$ by first-principles simulations.

We focus on the effect of polarization reorientation on the $r^\eta_{xx,x}$ electro-optic coefficient. As shown in Fig.~\ref{fig4}(a), the $x$ direction in $r^\eta_{xx,x}$ is defined with respect to the global $xyz$ coordinate system, and the electric field is rotated from the $[111]$ direction to the $[\bar{1}11]$ direction within the gray plane. The magnitude of the electric field is fixed at 0.3 MV/cm. In this case, the upper bounds of $\theta$ and $\phi$ angles are $\sim$71$^\circ$. Similar to SnS and SnSe, BaTiO$_3$ and KNbO$_3$ bulk ferroelectrics also exhibit hysteretic behaviors between $\theta$ and $\phi$ angles [see Fig.~\ref{fig4}(b)]. The electro-optic $r^\eta_{xx,x}$ coefficients as functions of $\theta$ are shown in Fig.~\ref{fig4}(c). When changing the $\theta$ angle, the $r^\eta_{xx,x}$ coefficients for KNbO$_3$ and BaTiO$_3$ can be enhanced to $\sim$226 (at $\theta\approx 38^\circ$) and $\sim$140 pm/V (at $\theta\approx 55^\circ$), respectively. These values are much larger than $\sim$21 pm/V for KNbO$_3$ and $\sim$9 pm/V for BaTiO$_3$---the $r^\eta_{xx,x}$ coefficients obtained without applied electric fields. This indicates that polarization reorientation is also a valid strategy for enhancing the Pockels responses in bulk multiaxial ferroelectrics.\\

\textit{Structural softness versus large Pockels effect—}The electro-optic responses in SnS, SnSe, BaTiO$_3$, and KNbO$_3$ are dominantly contributed by ionic motions. In Section III of the Supplementary Material \cite{SM}, we analyze the zone-center phonon modes of these multiaxial ferroelectrics. Upon polarization reorientation, the optical phonon modes in SnSe, BaTiO$_3$, and KNbO$_3$ exhibit substantial softening. In particular, the vibrational frequency of an optical mode in SnSe approaches zero at $\theta\approx 49^\circ$, which corresponds to the divergence of the $r^{\eta,\mathrm{2D}}_{xx,x}$ coefficient [see Fig.~\ref{fig3}(a)]. Our analysis confirms that structural softness (induced by polarization reorientation) is a underlying mechanism for engineering large Pockels effect in multiaxial ferroelectrics. \\

\textit{Summary—}To summarize, we have demonstrated that the polarization in a multiaxial ferroelectric may be reoriented by an electric field, the azimuth angles of the polarization and the electric field exhibit a hysteretic relationship, and such a polarization reorientation can lead to large Pockels responses. This strategy can be employed to enhance the linear electro-optic coefficients of various multiaxial ferroelectrics. Particularly, we show that the polarization reorientation significantly increases the electro-optic coefficients of SnSe (from $\sim$57 to $\sim$436 pm/V) and KNbO$_3$ (from $\sim$21 to $\sim$226 pm/V) by factors of $\sim$7 and $\sim$10, respectively. In conclusion, the reorientation of ferroelectric polarization provides an entirely new route for engineering large Pockels effect, and our work emphasizes the special role of multiaxial ferroelectrics toward this goal.\\

\noindent
\textit{Acknowledgment—}This work is supported by the National Key Research and Development Program of China (Grant No.~2022YFA1402502), the National Natural Science Foundation of China (Grant No.\,12374092), the Natural Science Basic Research Program of Shaanxi (Grant No.\,2023-JC-YB-017), the Shaanxi Fundamental Science Research Project for Mathematics and Physics (Grant No.\,22JSQ013), the Open Project of Key Laboratory of Material Simulation Methods \& Software of Ministry of Education, Jilin University (Grant No.\,202604), the Fundamental Research Funds for the Central Universities, the “Young Talent Support Plan” of Xi'an Jiaotong University, and the Xiaomi Young Talents Program.\\

\noindent
\textit{Data availability—}The data are not publicly available. The data are available from the authors upon reasonable request.

\appendix


\bibliography{main_ref}

\end{document}